\documentclass{llncs}
\usepackage{tikz}
\usepackage{pgfplots}
\usepackage{booktabs}
\usepackage{amssymb,amsmath,amsfonts}
\usepackage{epsfig,graphics,graphicx}

\usepackage{cleveref}
\usepackage{array}
\usepackage{pgfplotstable}
\usepackage{todonotes}

\newcount\mycount

\newcommand{\ah}{aH-index}
\newcommand{\h}{H-index}

\newcolumntype{L}[1]{>{\raggedright\let\newline\\\arraybackslash\hspace{0pt}}m{#1}}
\newcolumntype{C}[1]{>{\centering\let\newline\\\arraybackslash\hspace{0pt}}m{#1}}
\newcolumntype{R}[1]{>{\raggedleft\let\newline\\\arraybackslash\hspace{0pt}}m{#1}}

\newcommand{\NoOfAuthors}{20}

\crefname{figure}{fig.}{figs.}

\begin{document}

\title{Author Evaluation Based on H-index and Citation Response}

\author{Milo\v{s} Kud\v{e}lka \and Jan Plato\v{s} \and Pavel Kr\"{o}mer}

\institute{
Department of Computer Science\\
V\v{S}B Technical University of Ostrava\\
Ostrava, Czech Republic\\
\email{\{milos.kudelka, jan.platos, pavel.kromer\}@vsb.cz}\\
}

\maketitle
\begin{abstract}
An accurate and fair assessment of the efficiency and impact of scientific work is, despite a lot of recent research effort, still an open problem. The measurement of quality and success of individual scientists and research groups can be approached from many different directions, none of which is universal. A reason for this is inherently different behavior of different scientists within the global research community. A complex evaluation of ones publication activities requires a careful consideration of a wide variety of factors. The well-known \h ~is one of the most used bibliometric indices. Despite its many imperfections, its simplicity and ease of interpretation make it a popular scientometric method.
This short paper uses the ideas behind the \h ~to analyze communities of authors who cite publishing scientists. A new author evaluation measure named \ah is proposed, and intuitive interpretations of its properties and semantics are presented. Preliminary experiments with authors with high \h ~active in the area of computer science are presented to demonstrate the properties of the proposed measure.        
\end{abstract}


\section{Introduction}
Publication activities and citation response of active scientists are affected by a number of different factors. Natural preconditions for successful publication of research results include a deep understanding of a research topic, creativity, diligence, and last but not least an ability to comprehend current trends and latest advances in a field of study.
Many new factors influencing the nature of a scientist's publication behavior can be attributed to modern technologies. Internet databases and search engines can be used to quickly explore information sources such as conference papers and journal articles. Electronic communication simplifies networking between scientists and their research groups and supports preservation of long and short-term contacts.     
Natural consequences include an increase in the average number of research paper co-authors, the emergence of strongly connected but sometimes opportunistic research groups (communities), and the rise of multidisciplinary research topics and hybrid methods. 

In response to this development, many new conferences and journals are launched every year. They are created in reaction to the wide need for the massive and rapid dissemination of research results. This need, however, is sometimes motivated by the academic and peer pressure exercised on individual scientists (publish or perish) rather than by the quality of their work.

A number of scientometric measures have been developed to evaluate ones research performance and impact of his or her work~\cite{martin1996use,rinia1998comparative}. The major issue of many simple author performance metrics, based on citation response, is that they often do not reflect the quality of research papers that triggered the response, and the publication venues (conferences and journals) that published them. Moreover, they do not reflect how the author reached his or her level of citation response. Important information such as with how many co-authors the scientist usually collaborates, how often and how much do different authors respond to his or her research in their work, and in how many research areas is he or she active, is not considered by most traditional scientometric measures at all.

Our approach, outlined in this short paper, is different. Even though we are very well aware of the known imperfections of the original \emph{\h}, we consider its underlying principles excellent and especially value its simplicity and ease of interpretation. In this work, we extend the \h ~by an analysis of the citation response received by scientists. The proposed evaluation measure, termed \emph{\ah}, complements the \h ~with a new type of information reflecting the quality and quantity of a scientist's citation response. Due to its design, heavily inspired by the principles of the \h, it suffers from the same problems. However, it also retains the simplicity and interpretability of the original Hirsch index.

%
%However, we want to discover more about publishing scientists. In general, we are most interested in the number and structure of authors and groups that that respond to the evaluated scientist in form of citations. For this purpose, a straightforward modification of the \emph{\h} is proposed. The proposed citation measure still suffers from similar problems as the original \emph{\h} but reveals a new type of information about the scientists. Moreover, it retains the simplicity and interpretability of the original Hirsch index. 

\section{Related Work}
Hirsch proposed in \cite{Hirsch15112005} a single number, \h, as a particularly simple and useful way to characterize the scientific output of a researcher. A purpose of the \h ~was to describe both the productivity and impact of the published work of a scientist. However, there are some well-known drawbacks of using the \h~to evaluate and compare individual scientists. They are e.g. neglecting the quality of publications, a number of co-authors of a citing paper, comparing scientists working in different research fields, the number of citations of most cited papers, etc. That is why \h~characteristics were extensively investigated. Costas et al. analyzed in \cite{costas2007h} the relationship of the \h~to other well-known bibliometric indicators.

After the introduction of the \h, many improvements that addressed its fundamental drawbacks were proposed. In \cite{zhang2009index}, Zhang proposed a new index that is suitable for evaluating highly cited scientists and comparing groups of scientists with an identical \h. Alonso et al.  present in \cite{Alonso2009h} a comprehensive review on the \h~and related indicators. They studied their main advantages, drawbacks, and main applications. In \cite{bornmann2011multilevel}, Bornmann et al. present a study of 37 different variants of \h. They show a high correlation between the \h~and most of its variants.

The complicated relationship between the scientist and his or her co-authors is one of the problems of the \h. Hirsch proposes in \cite{hirsch2010index} a new version of the \h~that takes into account the effect of multiple co-authors and solves a well-known problem with the so-called Hirsch core. A similar problem is solved and a new variant of the \h~is introduced by Wan et al. in \cite{wan2007pure}. In \cite{mccarty2013predicting}, McCarty et al. apply social network analysis on ego co-authorship network. They show that the highest \h~can be achieved by working with many co-authors.

Analysis of the relation between the \h~and the behavior of citing authors (citers) is also suitable for a better understanding how a community of citing authors influences \h. Brooks study complex citer motivations in \cite{brooks1986evidence}. Seven citer motives are analyzed, and more than 70\% of references surveyed are the result of a complex interplay of multiple citer motives. Amancio et al. investigate in \cite{amancio2012three} the dependency of a quantity of citations on author reputation (visibility). They show that the reputation can affect a temporal evolution of \h. In \cite{bras2011bibliometric}, Bras-Amor{\'o}s et al. present a new index in which the evaluated objects are the citations received by an author and the quality function is based on a collaboration distance between the authors of the cited and the citing papers. The new index takes into account only significant citations; significance is proportional to collaboration distance.

\section{Author Evaluation}
\label{sec:eval}
In this section, we first recall the \h ~and then propose a new citation measure evaluating certain \emph{properties} of the citation response received by publishing scientists. 
%Finally, we discuss its interpretation and propose several hypotheses about its properties. 

\begin{definition}[H-index]
A scientist has (Hirsch) index $\mathrm{h}$ if $\mathrm{h}$ of his or her $\mathrm{N_p}$ papers have at least $\mathrm{h}$ citations each and the other $\mathrm{(N_p - h)}$ papers have $\leq \mathrm{h}$ citations each~\cite{Hirsch15112005}. 
\vskip0.25em\noindent
All papers by a scientist that have at least $\mathrm{h}$ citations form his or her \emph{Hirsch core}~\cite{rousseau2006new}.
\end{definition}
%
%\subsection{\ah}
The proposed citation measure, called \ah, is designed to numerically evaluate the following intuition: the impact of a scientist on a research community is proportional to the number of his or her publications  that are cited by the members of the community. We are seeking for a measure reflecting how much and how often do different researchers cite the publications of the evaluated scientist.

\begin{definition}[\ah]
A scientist has \ah~ $\mathrm{a}$ if $\mathrm{a}$ of the $\mathrm{N_c}$ researchers, that cite his or her work, cite at least $\mathrm{a}$ his or her publications each and the other $\mathrm{(N_c - a)}$ researchers cite $\leq \mathrm{a}$ publications.
%\vskip0.25em\noindent
%All authors that cite at least $\mathrm{a}$ papers of a scientist form his or her \emph{aH source}. We call the authors in ones aH source \emph{significant followers}.
%\vskip0.25em\noindent
%The set of all publications that are cited by ones significant followers is called \emph{aH core}.
\end{definition}
The \ah ~comprehends certain qualitative aspects of the citation response. A high \ah ~is awarded to scientists who exhibit prolific publication activities (i.e. they produce a high number of papers) and at the same time achieve a significant impact on the research community (i.e. a high number of other researchers cite a high number of their publications). Such a high \ah ~indicates that there is a large group of other researchers that follow the work of the evaluated scientist and cite many of his or her publications.

Nevertheless, a high value of the \ah ~can be also obtained by an extremal behavior, contradicting the intuition mentioned above. Let us assume that a group of 20 authors writes a single research paper in which they cite 20 different publications of a single scientist. Such a scientist is immediately awarded 20 points of \ah ~due to this research paper alone. The impact of such publication on the scientific community as a whole is, however, questionable (it has been referred to in a single work only). The likelihood of such extremal behavior in current bibliographic datasets is a subject of our ongoing investigation.
% 
%An example may be Alessandro F. Garcia with \emph{\ah} ~34 and size of aH core 84. 
%This vulnerability to extremal citation behaviour is clearly a weak spot of the \emph{\ah} in the present form.

\section{Experiments}
The properties of the \ah ~is initially investigated using a real-world bibliographical dataset. The dataset was aggregated under the Arnet Miner project\cite{Tang:08KDD} and is publicly available on its website\footnote{http://aminer.org/big-scholar-challenge/}. It consists of 2,092,209 papers, 8,024,869 citations between papers, and 1,712,433 authors with related information. 
%We also use the supplementary information that maps authors to their papers. 
This dataset is large enough to allow deep analysis but suffers from several problems, e.g. missing author from a publication, author duplication and misspelling, etc.
%Co-authors of some publications are missing (i.e. not included in the list of authors) or are not linked to their papers correctly. Moreover, the list of authors is not normalized so one author can appear under multiple identities due to, for example, spelling errors. Specifically the authors with accented letters in their names are often present several times. 
Nevertheless, it is a unique real-world bibliographic dataset that can be used to validate properties of scientometric indicators.
After pre-processing, we used a subset of dataset containing $14,744$ scientist with an \h ~of at least $8$ for our initial computational experiments. The results of this dataset are used to formulate several hypotheses about the proposed \ah. The confirmation of these hypotheses requires a thorough statistical analysis of detailed author data that is beyond the scope of this short paper. A thorough validation of the outlined concepts and propositions is the subject of our current research.  

\begin{proposition}
If the \ah ~is taken as a complement to the original \h, it can uncover and quantify certain publication behavior that a scientist used to attain his or her \h. It can be also used to evaluate how much and how strongly (persistently) he or she influenced other researchers in the community.
\end{proposition}
\begin{figure}[h!]
%\begin{figure}[h!]
 \begin{center}
 	\includegraphics[width=12cm]{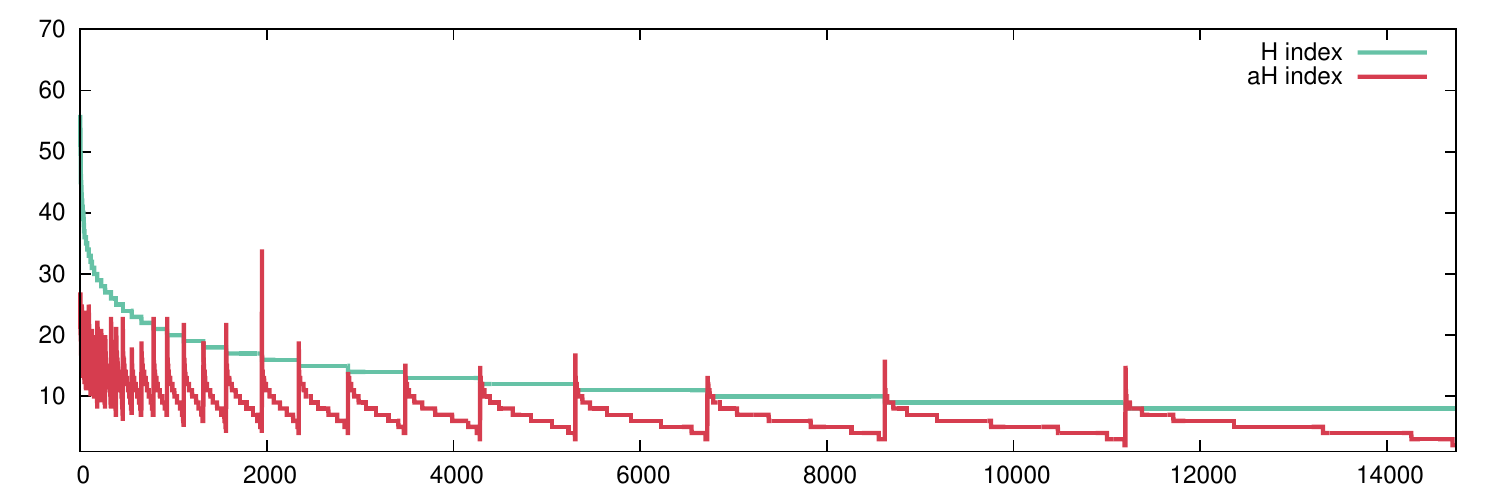}
 \caption{The distribution of \h ~and \ah ~in test data.}
 \label{fig:1}
 \end{center}
\vspace{-7mm}
\end{figure}

\Cref{fig:1} compares the distributions of the \h ~and the proposed \ah, respectively. It clearly shows that the same value of the \h ~can be achieved with both very high and very low values of the \ah. In other words, the \ah ~provides a more detailed and more sensitive assessment of publication activities with respect to received citation response. 

The implications of the \ah ~for the evaluation of scientists are, however, subject to further investigation. One possible interpretation of the effect of a high \ah ~is that achieving a high \h ~is easier when one has a high \ah. Our motivation for this hypothesis is the following. If there are not many citers that would cite a large number of a scientist's publications (i.e., he or she has low \ah), but if the scientist has a high \h, there had to be a wide group of citers that respond to only a few of his publications each. If there is a group of citers that cite a high number of his or her publications, they all contribute to the \h ~and attaining a high $h$ is easier.  
In our work, we expect that very high values of \ah, close to the values of the \h, indicate an abnormal behavior of citers. The spikes in the value of \ah ~on~\cref{fig:1} correspond to such abnormal situations. We also consider a combination of low \ah ~and high \h ~a sign of abnormal behavior of citers. 

As shown in~\cref{fig:1}, the majority of authors in the test dataset has a higher \h ~than \ah. The arithmetic mean of the ratio of the \h, $h$, and \ah, $a$, in this dataset for authors with $h \geq 8$ is approximately $1.848$. We use this mean as a correction coefficient $r$, and evaluate its relation to $h$.

Let $n = \frac{h}{r}$ be a normal value of the \ah. The value of $n$ for each scientist with \h ~$h$ tells, what should his or her \ah ~$a$ be if he or she received a citation response typical (average) for a given community. Next, for each scientist consider an \emph{xA-ratio}, $x$, as a proximity (similarity) of \ah, $a$,  to the normal value $n$:
\begin{align}
\label{eq:1}
x = \begin{cases}
\frac{a}{n}, & \text{if } a \leq n\\
\frac{n}{a}, & \text{if } a > n\\
\end{cases}.
\end{align} 
In the \cref{eq:1}, we expect that each evaluated scientist has at least one publication and at least one citation. We can now use the \emph{xA-ratio} to modify the \h. Their product, $x\cdot h$, can be interpreted as a correction of the \h ~with respect to the behavior of citers.
For authors with typical publication activities and receiving citation response typical for given community, this product should not introduce large changes to $h$. 
Rather, it provides a subtle correction to the \h ~that applies a penalty for the following abnormal types of citation response:
\begin{enumerate}
  \item citers respond to an above average number of scientist's publications,
  \item citers respond to a bellow average number of scientist's publications.
\end{enumerate}     

As described earlier, an above average citation response can also be obtained by a single publication in which a large number of authors cite a large number of a scientist's papers. Such publications can be understood as \emph{citation bombs} and it is a question whether they should be omitted when computing $a$ (and eventually $h$).  
\Cref{tab:tabAAA} shows the top {\NoOfAuthors} scientists in the test dataset ordered by the \h ~and $x\cdot h$, respectively. In this experiment, we expect that the value of $h$ should be $1.848$ times higher than $a$. As previously mentioned, this was the average ratio of $h$ and $a$ in the test dataset. %The correction coefficient, $r$, was set to $1.848$.

We can use the discussed measures to identify scientists with a high $h$ and at the same time an exceptional, i.e. above or bellow average, $a$. The former case can indicate that such scientist is favored by a community of citers with an outstanding behavior characterized by a tendency to cite a large number of ones publications. An example of in the test dataset is Alessandro F. Garcia with $h = 16$ and $a = 34$. The latter case indicates a very wide impact of the publications of the evaluated scientist, e.g. David R. Karger with $h = 36$ and $a = 11$.   

In the next experiments, we set the correction coefficient, $r$, to the minimum and maximum value in our dataset. The minimum value corresponds to $h=16, a=34$ and the maximum value to $h=18, a=4$ (Richard A. Caruana). For the minimum value we expect that $a$ is higher than $h$, the maximum value assumes that $h$ is more than four times higher than $a$.
%\Cref{tab:tabBBB} shows the top {\NoOfAuthors} scientists ordered by $x\cdot h$ for both values of $r$.    
\Cref{tab:tabBBB} shows that even after correction, most of the top {\NoOfAuthors} authors feature a higher \h.

The correlation between $x\cdot h$ and $h$  in the test dataset for 14,744 scientists with $h \geq 8$ and correction coefficient $r = 1.848$ is equal to $0.937$. If we focus on the top scientists in the test dataset, we can see that the correlation between $x\cdot h$ and $h$ is 0.839 for 1111 scientists with $h \geq 20$ and 0.720 for 184 scientists with $h \geq 30$. It means that despite being based on the \h, the corrected evaluations produce different rankings of authors and provide different assessments of a scientist's publication activities.

\begin{table}%
\caption{The first {\NoOfAuthors} authors according the H-index and with reduction for $r=1.848$}
\label{tab:tabAAA}
\centering
\begin{filecontents*}{tab_AAA.dat}
Name	H-index	AH-index	Name	H-index	AH-index	H-REDUCTION
Hector Garcia-Molina	60	19	Jiawei Han	53	27	49.896
Scott Shenker	56	21	David E. Culler	51	25	46.2
Jiawei Han	53	27	Chris Faloutsos	50	24	44.352
David E. Culler	51	25	Ian Foster	46	24	44.352
Chris Faloutsos	50	24	Anil K. Jain	50	22	40.656
Anil K. Jain	50	22	P S Yu	46	22	40.656
Jeffrey D. Ullman	49	19	Mihir Bellare	42	24	39.77272727
Ian Foster	46	24	Scott Shenker	56	21	38.808
P S Yu	46	22	W Bruce Croft	44	21	38.808
D Estrin	45	20	Moshe Y. Vardi	42	21	38.808
Jennifer Widom	45	19	M. Naor	42	21	38.808
Ch. H. Papadimitriou	45	17	David J. DeWitt	41	21	38.808
Hari Balakrishnan	45	17	Tom Henzinger	42	25	38.18181818
Jon M. Kleinberg	45	15	E. M. Clarke	40	23	37.6435159
W Bruce Croft	44	21	HongJiang Zhang	39	20	36.96
Rakesh Agrawal	43	19	D Estrin	45	20	36.96
Ben Shneiderman	43	18	I. Stoica	41	20	36.96
T. Anderson	43	17	Dan Suciu	40	20	36.96
Mihir Bellare	42	24	Serge Abiteboul	37	20	36.96
Moshe Y. Vardi	42	21	Dan Boneh	37	20	36.96
M. Naor	42	21	Oded Goldreich	40	24	36.07503608
Tom Henzinger	42	25	Robert Endre Tarjan	39	19	35.112
Michael Stonebraker	42	19	Hector Garcia-Molina	60	19	35.112
Pat Hanrahan	42	18	Jennifer Widom	45	19	35.112
David J. DeWitt	41	21	Rakesh Agrawal	43	19	35.112
I. Stoica	41	20	Michael Stonebraker	42	19	35.112
Ronald Fagin	41	16	W. Y. Ma	40	19	35.112
David A. Patterson	41	15	Rajeev Alur	36	19	35.112
E. M. Clarke	40	23	Amir Pnueli	36	19	35.112
Dan Suciu	40	20	William Buxton	36	19	35.112
Oded Goldreich	40	24	Jeffrey D. Ullman	49	19	35.112
W. Y. Ma	40	19	Thomas W. Reps	35	19	34.88835726
Joseph M. Hellerstein	40	18	Ken Kennedy	36	21	33.39517625
R Motwani	40	17	Pat Hanrahan	42	18	33.264
Randy Katz	40	13	Joseph M. Hellerstein	40	18	33.264
HongJiang Zhang	39	20	Andrew Zisserman	39	18	33.264
Robert Endre Tarjan	39	19	Vern Paxson	39	18	33.264
Andrew Zisserman	39	18	R. Szeliski	35	18	33.264
Vern Paxson	39	18	S Osher	35	18	33.264
Raghu Ramakrishnan	39	17	Surajit Chaudhuri	34	18	33.264
P Raghavan	38	15	Ben Shneiderman	43	18	33.264
Monica S. Lam	38	15	Leo J. Guibas	35	20	33.14393939
Serge Abiteboul	37	20	A. Gupta	33	18	32.73809524
Dan Boneh	37	20	Ravin Balakrishnan	35	21	31.56565657
M. Frans Kaashoek	37	16	Ch. H. Papadimitriou	45	17	31.416
M Abadi	37	15	Hari Balakrishnan	45	17	31.416
Mihalis Yannakakis	37	13	R Motwani	40	17	31.416
L Zhang	37	12	Raghu Ramakrishnan	39	17	31.416
Rajeev Alur	36	19	Jeff Naughton	35	17	31.416
Amir Pnueli	36	19	Gerhard Weikum	34	17	31.416

\end{filecontents*}
\pgfplotstabletypeset[
every head row/.style={after row={\hline}},
row predicate/.code={\ifnum#1<\NoOfAuthors \relax \else \pgfplotstableuserowfalse \fi},
col sep=tab,
columns/Name/.style={string type,column type=l,column name={Name}},
columns/{H-index}/.style={column type=C{7mm},column name={h}},
columns/{AH-index}/.style={column type=C{7mm},column name={aH}},
columns/{Name--index3}/.style={string type,column type=|l,column name={Name}},
columns/{H-index--index4}/.style={column type=C{7mm},column name={h}},
columns/{AH-index--index5}/.style={column type=C{7mm},column name={aH}},
columns/{H-REDUCTION}/.style={column type=C{12mm},fixed zerofill,precision=2,column name={r=1.848}},
]{tab_AAA.dat}
\end{table}
\begin{table}%
\caption{The first {\NoOfAuthors} authors according the reduced H-index for $r = \frac{16}{34}$ and $r = \frac{18}{4}$}
\label{tab:tabBBB}
\centering
\begin{filecontents*}{tab_BBB.dat}
Name	H-index	AH-index	H-REDUCTION	Name	H-index	AH-index	H-REDUCTION
Alessandro F. Garcia	16	34	16	Hector Garcia-Molina	60	19	42.10526316
Jiawei Han	53	27	12.70588235	Scott Shenker	56	21	33.18518519
David E. Culler	51	25	11.76470588	Jon M. Kleinberg	45	15	30
Tom Henzinger	42	25	11.76470588	Jeffrey D. Ullman	49	19	28.08187135
W. van der Aalst	33	25	11.76470588	Randy Katz	40	13	27.35042735
Chris Faloutsos	50	24	11.29411765	Ch. H. Papadimitriou	45	17	26.47058824
Ian Foster	46	24	11.29411765	Hari Balakrishnan	45	17	26.47058824
Mihir Bellare	42	24	11.29411765	David R. Karger	36	11	26.18181818
Oded Goldreich	40	24	11.29411765	L Zhang	37	12	25.35185185
N. R. Jennings	36	24	11.29411765	Anil K. Jain	50	22	25.25252525
Jack J. Dongarra	33	23	10.82352941	Richard M. Karp	30	8	25
Milind Tambe	26	23	10.82352941	David A. Patterson	41	15	24.9037037
Micha Sharir	24	23	10.82352941	David Wagner	33	10	24.2
E. M. Clarke	40	23	10.82352941	T. Anderson	43	17	24.16993464
Brad A. Myers	36	23	10.82352941	Jennifer Widom	45	19	23.68421053
M. Harman	21	23	10.82352941	Mihalis Yannakakis	37	13	23.4017094
Francky Catthoor	20	23	10.82352941	Ronald Fagin	41	16	23.34722222
Anil K. Jain	50	22	10.35294118	Chris Faloutsos	50	24	23.14814815
Luca Benini	33	22	10.35294118	Richard Lipton	25	6	23.14814815
Silvio Micali	33	22	10.35294118	David E. Culler	51	25	23.12
J. Cong	29	22	10.35294118	Jiawei Han	53	27	23.11934156
Thomas Eiter	29	22	10.35294118	Ben Shneiderman	43	18	22.82716049
Feng Ding	17	22	10.35294118	Michael I. Jordan	35	12	22.68518519
P S Yu	46	22	10.35294118	D Estrin	45	20	22.5
Gonzalo Navarro	26	22	10.35294118	D S Weld	36	13	22.15384615
M. M. Burnett	19	22	10.35294118	Johannes E. Gehrke	36	13	22.15384615
Scott Shenker	56	21	9.882352941	Pat Hanrahan	42	18	21.77777778
W Bruce Croft	44	21	9.882352941	Tim Finin	28	8	21.77777778
Moshe Y. Vardi	42	21	9.882352941	Rakesh Agrawal	43	19	21.62573099
M. Naor	42	21	9.882352941	Donald E. Knuth	26	7	21.46031746
David J. DeWitt	41	21	9.882352941	Pattie Maes	26	7	21.46031746
Ken Kennedy	36	21	9.882352941	Daphne Koller	34	12	21.40740741
Ravin Balakrishnan	35	21	9.882352941	Don Towsley	34	12	21.40740741
Ronald P. Fedkiw	31	21	9.882352941	P Raghavan	38	15	21.39259259
Yufei Tao	29	21	9.882352941	Monica S. Lam	38	15	21.39259259
H. P. Seidel	29	21	9.882352941	P S Yu	46	22	21.37373737
Brent R. Waters	29	21	9.882352941	Ian H. Witten	31	10	21.35555556
John Mylopoulos	28	21	9.882352941	Ronald L. Rivest	31	10	21.35555556
A. B. Kahng	26	21	9.882352941	T Kanade	35	13	20.94017094
J Meseguer	25	21	9.882352941	R Motwani	40	17	20.91503268
Gheorghe Păun	25	21	9.882352941	N. H. Vaidya	29	9	20.7654321
HongJiang Zhang	39	20	9.411764706	Michael Stonebraker	42	19	20.63157895
D Estrin	45	20	9.411764706	W Bruce Croft	44	21	20.48677249
I. Stoica	41	20	9.411764706	M Abadi	37	15	20.28148148
Dan Suciu	40	20	9.411764706	M. Hearst	27	8	20.25
Leo J. Guibas	35	20	9.411764706	Eric A. Brewer	33	12	20.16666667
Hans-Peter Kriegel	34	20	9.411764706	Raghu Ramakrishnan	39	17	19.88235294
Elisa Bertino	33	20	9.411764706	Amin Vahdat	34	13	19.76068376
Georg Gottlob	31	20	9.411764706	B Liskov	34	13	19.76068376
Maurice Herlihy	30	20	9.411764706	Joseph M. Hellerstein	40	18	19.75308642
\end{filecontents*}
\pgfplotstabletypeset[
every head row/.style={after row={\hline}},
row predicate/.code={\ifnum#1<\NoOfAuthors \relax \else \pgfplotstableuserowfalse \fi},
col sep=tab,
columns/Name/.style={string type,column type=l,column name={Name}},
columns/{H-index}/.style={column type=C{7mm},column name={h}},
columns/{AH-index}/.style={column type=C{7mm},column name={aH}},
columns/H-REDUCTION/.style={column type=C{12mm},fixed zerofill,precision=2,column name={$r = \frac{16}{34}$}},
columns/{Name--index4}/.style={string type,column type=|l,column name={Name}},
columns/{H-index--index5}/.style={column type=C{7mm},column name={h}},
columns/{AH-index--index6}/.style={column type=C{7mm},column name={aH}},
columns/H-REDUCTION--index7/.style={column type=C{12mm},fixed zerofill,precision=2,column name={$r = \frac{18}{4}$}},
]{tab_BBB.dat}
\end{table}

\section{Conclusions}
\label{sec:conclusions}
This short paper presents a novel author evaluation measure based on the principles of the well-known \h ~and citation response. It outlines the principles of a new citation measure evaluating the behavior of groups of citers responding to scientist's publications. The new measure, called the \ah, was conceived as a complement to the original \h ~and allows the extraction of a new type of information about the evaluated scientists. 
The properties of this measure were studied using a comprehensive real-world bibliometric dataset. Based on the results of these initial experiments, the role and interpretation of the \ah ~was formulated and discussed. A notion of the \h ~modified by an \ah-based coefficient evaluating the citation response received by individual scientists, with respect to typical citation patterns in their community, was investigated.  

The work, summarized in this short paper, is indeed preliminary. However, it clearly outlines a number of promising options for a fair and accurate assessment of publication activities combining the best-of-breed scientometric methods with a novel idea coming in part from the area of network science.

%\section*{Acknowledgement}
%This work was supported  by grants of SGS, V\v{S}B - Technical University of Ostrava, no. SP2015/146. 

\bibliographystyle{splncs}
\bibliography{local}

\end{document}